\documentclass[twocolumn,showpacs,amsmath,amssymb,prc]{revtex4}
\usepackage{graphicx}
\usepackage{dcolumn}
\usepackage{bm}

\begin{document}

\title{Theoretical Standard Model Rates of Proton to Neutron \\ 
Conversions Near Metallic Hydride Surfaces}

\author{A. Widom}
\affiliation{Physics Department, Northeastern University, Boston MA 02115}
\author{L. Larsen}
\affiliation{Lattice Energy LLC, 175 North Harbor Drive, Chicago IL 60601}

\begin{abstract}
The process of radiation induced electron capture by protons or deuterons producing new 
ultra low momentum neutrons and neutrinos may be theoretically described within the standard 
field theoretical model of electroweak interactions. For protons or deuterons  in the 
neighborhoods of surfaces of condensed matter metallic hydride cathodes, such conversions are 
determined in part by the collective plasma modes of the participating charged particles, 
e.g. electrons and protons or deuterons. The radiation energy required for such low energy nuclear 
reactions may be supplied by the applied voltage required to push a strong charged current across a 
metallic hydride surface employed as a cathode within a chemical cell. The electroweak rates 
of the resulting ultra low momentum neutron production are computed from these considerations.  
\end{abstract}


\maketitle

\section{Introduction \label{intro}}

Excess heats of reaction have often been observed to be generated in  
the metallic hydride cathodes of certain electrolytic chemical cells. 
The conditions required for such observations include high electronic current 
densities passing through the cathode surface as well as high packing fractions 
of hydrogen or deuterium atoms within the metal. Also directly observed in 
such chemical cells are nuclear transmutations into elements {\em not} 
originally present prior to running a current through  and/or prior to 
applying a LASER light beam incident to the cathode 
surface\cite{Iwamura:1998,Violante:2002,Dash:2002,Miley:2005,Miley:1996,Miley:1997}. 
It seems {\em unlikely} that the direct cold fusion of two deuterons can be a 
requirement to explain at least many of such observations\cite{Pons:1989} because 
in many of these experiments, deuterons were initially absent.
For simplicity of presentation, we consider ``light water'' chemical cells 
in which deuterons are not to any appreciable degree present before the 
occurrence of heat producing nuclear transmutations.  

Nuclear transmutations in the work which follows are attributed to the creation and 
absorption of ultra low momentum neutrons as well as related production of neutrinos.  
Although other workers\cite{Mizuno:1998,Kozima:1998} have previously speculated on a 
central role for neutrons in such transmutations, they were unable to articulate a 
physically plausible mechanism that could explain high rates of neutron production 
under the stated experimental conditions. By contrast, in this work electrons are 
captured by protons all located in collectively oscillating "patches" on the metallic 
surface. Electrons are captured by protons all located in collectively oscillating 
``patches'' on the metallic surface. Since the energy threshold for such a reaction is 
\begin{equation}
Q_{in}\approx 
\left\{M_{\rm n}-(M_{\rm p}+m)\right\}c^2 
\approx 0.78233\ {\rm MeV}, 
\label{intro1}
\end{equation}
one requires a significant amount of initial collective radiation energy to 
induce the proton into neutron conversion  
\begin{equation}
{\rm (radiation\ energy)}+e^- +p^+ \to n+\nu_e.
\label{intro2}
\end{equation}
The radiation energy may be present at least in part due the power absorbed 
at the surface of the cathode. If \begin{math} {\cal V}  \end{math} denotes 
the voltage difference between the metallic hydride and the electrolyte and 
if \begin{math} {\cal J}  \end{math} denotes the electrical current per unit 
area into the cathode from the electrolyte, then 
the power per unit cathode surface area dissipated into infrared heat 
radiation is evidently 
\begin{equation}
{\cal P}={\cal V}{\cal J}=e{\cal V}\tilde{\Phi },
\label{intro3}
\end{equation}
wherein \begin{math} \tilde{\Phi } \end{math} is the flux per unit area of electrons 
exiting the cathode into the electrolyte. Typical metallic hydride cathodes will 
exhibit soft surface photon radiation in much the same physical manner as a ``hot wire'' 
in a light bulb radiates light. For the case of chemical cell cathodes, there will be 
a frequency {\em upward conversion} from virtually DC cathode currents and voltages up 
to infrared frequency radiation. Such an upward frequency conversion requires  
high order electromagnetic interactions between electrons, protons and photons.  

\begin{figure}[bp]
\scalebox {0.5}{\includegraphics{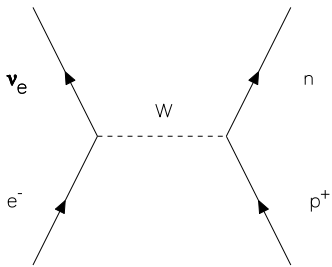}}
\caption{A low order diagram for   
$e^- +p^+ \to n+\nu_e$ in the vacuum is exhibited. 
In condensed matter metallic hydrides, 
the amplitude must include radiative corrections to very 
high order in $\alpha $.}
\label{fig1}
\end{figure}

The purpose of this work is to estimate the total rates of the reaction 
Eq.(\ref{intro2}) in certain metallic hydride cathodes. The lowest order 
vacuum Feynman diagram for the proton to neutron conversion  
is shown in FIG. \ref{fig1}. For the case of the reactions in metallic 
hydrides, one must include radiative corrections to FIG. \ref{fig1} to 
very high order in the quantum electrodynamic coupling strength  
\begin{equation}
\alpha =\frac{e^2}{4\pi \hbar c}\approx 
7.2973526 \times 10^{-3}.    
\label{intro4}
\end{equation}
The {\it W}-coupling in terms of the weak rotation angle 
\begin{math} \theta_W \end{math} will be taken to lowest order in 
\begin{equation}
\alpha_W =\frac{g^2}{4\pi \hbar c}=
\frac{\alpha }{\sin^2 \theta_W}\ .    
\label{intro5}
\end{equation}
Charge conversion reactions are weak due to the large mass 
\begin{math} M_W  \end{math} of the \begin{math} W^\pm \end{math}. 
The Fermi interaction constant, scaled by either the proton or 
electron masses, is determined\cite{Marshak:1969,Pokorski:2000} by    
\begin{eqnarray} 
G_F\approx \frac{\pi \alpha_W}{\sqrt{2}}\left(\frac{\hbar c}{M_W^2}\right),
\nonumber \\ 
\frac{G_FM_p^2}{\hbar c}\approx 1.02682\times 10^{-5},
\nonumber \\ 
\frac{G_Fm^2}{\hbar c}\approx  3.04563\times 10^{-12}.    
\label{intro6}
\end{eqnarray}  
In the work which follows, it will be shown how the weak proton 
to ultra low  momentum neutron conversions on metallic hydride surfaces 
may proceed at appreciable rates in spite of the small size of the Fermi weak 
coupling strength. 

An order of magnitude estimate can already be derived from a four 
fermion weak interaction model presuming a previously 
discussed\cite{Widom:2006} electron mass renormalization 
\begin{math} m\to \tilde{m}=\beta m  \end{math} due 
to strong local radiation fields. Surface electromagnetic modes excited 
by large cathode currents can add energy to a bare electron state 
\begin{math} e^- \end{math} yielding a mass renormalized heavy electron 
state \begin{math} \tilde{e}^- \end{math}, with  
\begin{equation} 
\tilde{m}=\beta m.
\label{est1} 
\end{equation}
The threshold value for the renormalized electron mass which allows 
for the reaction Eqs.(\ref{intro1}) and (\ref{intro2}) is   
\begin{equation}
\beta > \beta_0\approx 2.531.
\label{est2}
\end{equation}
For a given heavy electron-proton pair 
\begin{math} (\tilde{e}^- p^+) \end{math}, 
the transition rate into a neutron-neutrino pair may be estimated in the 
Fermi theory by  
\begin{eqnarray}
\Gamma_{(\tilde{e}^- p^+)\to n+\nu_e }\sim 
\left(\frac{G_Fm^2}{\hbar c}\right)^2\left(\frac{mc^2}{\hbar }\right)
(\beta-\beta_0)^2,
\nonumber \\ 
\Gamma_{(\tilde{e}^- p^+)\to n+\nu_e }\sim 
9\times 10^{-24}\left(\frac{mc^2}{\hbar }\right)
(\beta-\beta_0)^2,
\nonumber \\ 
\Gamma_{(\tilde{e}^- p^+)\to n+\nu_e }\sim 
7\times 10^{-4}\ {\rm Hz}\times (\beta-\beta_0)^2,
\label{est3}
\end{eqnarray}
If there are \begin{math} n_2 \sim 10^{16}/{\rm cm^2} \end{math} 
such \begin{math} (\tilde{e}^- p^+) \end{math} pairs per unit 
surface area within several atomic layers below the cathode surface, 
then the neutron production rate per unit surface area per unit time 
may be estimated by 
\begin{eqnarray} 
\varpi_2 \approx n_2\Gamma_{(\tilde{e}^- p^+)\to n+\nu_e }\ ,
\nonumber \\ 
\varpi_2\sim \left(\frac{10^{13}\ {\rm Hz}}{{\rm cm}^2}\right)
\times (\beta-\beta_0)^2 .
\label{est4}
\end{eqnarray}
Significantly above threshold, say 
\begin{math} \beta \sim 2\beta_0\sim 5 \end{math},  
the estimated rate 
\begin{math} \varpi_2\sim 10^{13}\ {\rm Hz}/{\rm cm}^2 \end{math} 
is sufficiently large so as to explain observed nuclear transmutations in 
chemical cells in terms of weak interaction transitions of 
\begin{math} (\tilde{e}^- p^+) \end{math} 
pairs into pairs into neutrons and neutrinos and the subsequent absorption 
of these ultra low momentum neutrons by local nuclei.

It is worthy of note that the total cross section for the scattering of neutrons 
with momentum \begin{math} p  \end{math} may be written 
\begin{math} \sigma_{tot}=(4\pi \hbar/p){\Im m}{\cal A}(p)\end{math} wherein the 
forward scattering amplitude for neutrons is 
\begin{math} {\cal A}(p) \end{math}\cite{Larsen:2005}. In the ultra-low momentum limit 
\begin{math} p\to 0 \end{math}, the cross section for neutron absorption associated 
with a complex scattering length 
\begin{math} b=\lim_{p\to 0} {\Im m}{\cal A}(p) \end{math} 
formally diverges. For a finite but large neutron wavelength 
\begin{math} \lambda  \end{math}, the mean free neutron absorption path 
length \begin{math} \Lambda  \end{math} corresponding to \begin{math} n_a \end{math} 
absorbers per unit volume obeys 
\begin{equation}
\Lambda ^{-1}=n_a\sigma_{tot}\approx 2n\lambda b. 
\label{Lambda1}
\end{equation}
Numerically, for \begin{math} n_a\sim 10^{22}/{\rm cm}^3  \end{math} neutron 
absorbers per unit volume with an imaginary part of the scattering length 
\begin{math} b\sim 10^{-13}\ {\rm cm}  \end{math}
and with ultra-low momentum neutrons formed with a wavelength of 
\begin{math} \lambda \sim 10^{-3}\ {\rm cm}  \end{math}, a neutron will move 
on a length scale of \begin{math} \Lambda \sim 10^{-6}\ {\rm cm} \end{math} 
before being absorbed. The externally detectible neutron flux into the laboratory 
from the cathode is thereby negligible\cite{Widom:2006}.

Similarly, there is a strong suppression 
of gamma-ray emission due to the absorption of such rays by the heavy surface 
electrons\cite{Larsen:2005}. The mean free path of a photon in a metallic 
condensed matter system is related to the conductivity 
\begin{math} \sigma  \end{math}; employing Maxwell's equations and the equations 
\begin{equation}
\Lambda_\gamma ^{-1}=\frac{\sigma }{c}=R_{vac}\sigma 
\approx 4\alpha \left(\frac{\pi }{3}\right)^{1/3}n^{2/3}\bar{l}
\label{Lambda2}
\end{equation}
wherein \begin{math} n \end{math} is the density per unit volume of heavy electrons 
on the cathode surface and \begin{math} \bar{l} \end{math} is the heavy electron 
mean free path. For hard photon energies in the range 
\begin{math} 0.5\ {\rm Megavolt}<(\hbar \omega_{\gamma }/e) <10\ {\rm Megavolt}  \end{math} 
we estimate \begin{math} n^{2/3}\sim 10^{15}/{\rm cm}^2 \end{math} and 
\begin{math} \bar{l}\sim 10^{-6}\ {\rm cm} \end{math} leading to the short gamma-ray 
mean free path \begin{math} L_\gamma \sim 3.4\times 10^{-8}\ {\rm cm}  \end{math}. 
In estimating such a small mean free path for hard gamma photon, let us remind the reader 
of the inadequate nature of single electron single photon scattering when making estimates
of photon mean free paths. For example, a single optical photon scatters off a single 
single non-relativistic electron with a Thompson cross section. For many electrons, 
if the Thompson cross section were applied then one might conclude that metals were transparent 
in the optical regime. The proper estimates of photon absorption rates {\em in condensed matter}  
employs the electrical conductivity.

In Sec.\ref{NS}, an exact expression is derived for the emission  
rate \begin{math} \varpi  \end{math} per unit time per unit volume for 
creating neutrinos. It is then argued, purely on the basis of 
conservation laws, that \begin{math} \varpi  \end{math} also represents 
the rate per unit time per unit volume of neutron production.
The rate \begin{math} \varpi  \end{math}, in Sec.\ref{CCF}, is expressed 
in terms of composite fields consisting of charged electrons and opposite  
charged {\it W}-bosons. The effective {\it W}-bosons for condensed matter 
systems may be written to a sufficient degree of accuracy in terms of 
Fermi weak interaction currents 
\begin{eqnarray}
{\cal I}^+_\mu =c\left(\bar{\psi}_n  \gamma_\mu (g_V-g_A\gamma_5)\psi_p\right),
\nonumber \\ 
{\cal I}^-_\mu =c\left(\bar{\psi}_p  \gamma_\mu (g_V-g_A\gamma_5)\psi_n\right), 
\label{intro7}
\end{eqnarray} 
wherein the Dirac matrices are defined in Sec.\ref{NS},  
\begin{math} \psi_p \end{math} and \begin{math} \psi_n \end{math} 
represent, respectively, the proton and neutron Dirac fields and the 
vector and axial vector coupling strengths are determined by 
\begin{eqnarray}
\lambda \equiv \frac{g_A}{g_V}\approx 1.2695,
\nonumber \\ 
\cos \theta_C\equiv g_V\approx 0.9742, 
\label{intro8}
\end{eqnarray} 
wherein \begin{math} \theta_C  \end{math} is a strong interaction 
quark rotation angle. In Sec.\ref{EMR}, the electron fields as  
renormalized by metallic hydride surface radiation are explored and   
the effective mass renormalization in Eq.(\ref{est1}) is 
established. In Sec.\ref{EPO}, we consider the coupled electron and 
proton oscillations near the surface of a metallic hydride.

\begin{figure}[tp]
\scalebox {0.6}{\includegraphics{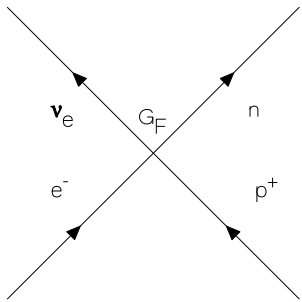}}
\caption{The four fermion vertex for     
$e^- +p^+ \to n+\nu_e$ in the vacuum is exhibited. 
In the large $M_W$ limit, the Feynman diagram of 
FIG. \ref{fig1} collapses into the above Feynman diagram.
In condensed matter metallic hydrides, the resulting 
effective $W^\pm $ fields are defined in 
Eqs.(\ref{intro7}) and (\ref{CCF1W}).}
\label{fig2}
\end{figure}

In Sec.\ref{ISPW} the nature of the neutron production is 
discussed in terms of {\em isotopic spin waves}. In the limit 
in which the protons and neutrons are {\em non-relativistic}, 
one may view the proton and neutron as different isotopic spin 
states of a nucleon\cite{Cassin:1936} with the charged proton 
having an isotopic spin \begin{math} +1/2  \end{math} and with 
the neutron having an isotopic spin \begin{math} -1/2  \end{math}. 
If \begin{math} n({\bf r}) \end{math} and 
\begin{math} p({\bf r}) \end{math} 
represent, respectively, the two (real) spin component fields 
for non-relativistic neutrons and protons, then the operator 
isotopic spin density 
\begin{math} 
{\bf T}({\bf r})=
\big({\cal T}_1({\bf r}),{\cal T}_2({\bf r}),{\cal T}_3({\bf r}) \big) 
\end{math} of 
the many body neutron-proton states may be written 
\begin{eqnarray}
{\cal T}_1({\bf r})&=&\frac{1}{2}
\left(p^\dagger ({\bf r})n({\bf r})
+n^\dagger ({\bf r})p({\bf r})\right),
\nonumber \\ 
{\cal T}_2({\bf r})&=&\frac{i}{2}
\left(n^\dagger ({\bf r})p({\bf r})
-p^\dagger ({\bf r})n({\bf r})\right),
\nonumber \\ 
{\cal T}_3({\bf r})&=&\frac{1}{2}
\left(p^\dagger ({\bf r})p({\bf r})
-n^\dagger ({\bf r})n({\bf r})\right),
\nonumber \\ 
{\cal T}^\pm({\bf r})&=&T_1({\bf r})\pm iT_2({\bf r}).
\label{intro9}
\end{eqnarray}
In the non-relativistic limit, these isotopic spin operators determine 
the time-component of Fermi weak interaction currents in 
Eq.(\ref{intro7}) via 
\begin{equation}
{\cal I}^{\mp \ 0} ({\bf r})\approx cg_V{\cal T}^\pm({\bf r}).
\label{intro10}
\end{equation}
The remainder of the non-relativistic weak currents are of the 
Gamow-Teller variety\cite{Gamow:1936} and require the true spin as 
well as isotopic spin version of Eq.(\ref{intro10}); i.e. 
with \begin{math} {\bf S}={\bf \sigma}/2 \end{math} as the 
Fermion spin matrices, the combined 
spin and isotopic spin operator densities are
\begin{eqnarray}
{\cal S}_{1,j}({\bf r})&=&
\left(p^\dagger ({\bf r})S_j n({\bf r})
+n^\dagger ({\bf r})S_j p({\bf r})\right),
\nonumber \\ 
{\cal S}_{2,j}({\bf r})&=&i
\left(n^\dagger ({\bf r})S_jp({\bf r})
-p^\dagger ({\bf r})S_jn({\bf r})\right),
\nonumber \\ 
{\cal S}_{3,j}({\bf r})&=&
\left(p^\dagger ({\bf r})S_j p({\bf r})
-n^\dagger ({\bf r})S_j n({\bf r})\right),
\nonumber \\ 
{\cal S}^{\pm \ j}({\bf r})&=&{\cal S}_{1,j}({\bf r})
\pm i{\cal S}_{2,j}({\bf r}).
\label{intro11}
\end{eqnarray}
In the non-relativistic limit for the protons and neutrons, 
the spatial components of the weak interaction currents 
in Eq.(\ref{intro7}) are 
\begin{equation}
{\cal I}^{\mp \ j} ({\bf r})\approx 
-cg_A{\cal S}^\pm _j({\bf r}).
\label{intro12}
\end{equation}
Altogether, in the nucleon non-relativistic limit
\begin{equation}
{\cal I}^{\mp \ \mu} \approx 
c\left(-g_A{\cal S}^\pm _1,-g_A{\cal S}^\pm _2,
-g_A{\cal S}^\pm _3,g_V{\cal T}^\pm \right).
\label{intro13}
\end{equation}
The isotopic formalism describes the neutron creation 
as a surface isotopic spin wave.  
Out of many oscillating protons in a surface patch, 
only one of these protons will convert into a neutron. 
However, one must superimpose charge conversion 
amplitudes over all of the possibly converted protons 
in the patch. This describes an isotopic spin wave localized in the 
patch with wavelength \begin{math} k^{-1}  \end{math}. 
The wavelength in turn describes the ultra low momentum 
\begin{math} p\sim \hbar k  \end{math} of the produced 
neutron. Finally in the concluding Sec.\ref{conc}, further numerical  
estimates will be made concerning the weak interaction 
production rate of such neutrons.

\section{Neutrino Sources \label{NS}}

The conventions here employed are as follows: The Lorentz metric 
\begin{math} \eta^{\mu \nu} \end{math} has the signature 
\begin{math} (+,+,+,-) \end{math} so that the Dirac matrix algebra 
may be written 
\begin{equation}
\gamma^\mu \gamma^\nu =-\eta^{\mu \nu}-i\sigma^{\mu \nu}
\ \ \ {\rm wherein}\ \ \ \sigma^{\mu \nu}=-\sigma^{\nu \mu }.
\label{NS1}
\end{equation}
The chiral matrix \begin{math} \gamma_5  \end{math} is 
defined with the antisymmetric symbol signature 
\begin{math} \epsilon_{1230}=+1 \end{math} employing   
\begin{equation}
\frac{1}{4!}\epsilon_{\mu \nu \lambda \sigma }\gamma^\mu \gamma^\nu 
\gamma^\lambda \gamma^\sigma =i\gamma_5   
\label{NS2}
\end{equation}
and chiral projection matrices are thereby 
\begin{equation}
P_{\pm }=\frac{1}{2}\left(1\mp \gamma_5\right).
\label{NS3}
\end{equation} 
Further algebraic matrix identities of use in the work below, 
such as 
\begin{eqnarray}
\gamma^\lambda \gamma^\mu \gamma^\sigma P_\pm =
\pm h^{\lambda \mu \sigma \nu}\gamma_\nu P_\pm \ ,
\ \ \ \ \ \ \ \ \ \ 
\nonumber \\ 
h^{\lambda \mu \sigma \nu}=i\epsilon^{\lambda \mu \sigma \nu}
-\eta^{\lambda \mu }\eta^{\sigma \nu }
+ \eta^{\lambda \sigma }\eta^{\mu \nu }
-\eta^{\mu \sigma }\eta^{\lambda \nu } , 
\label{NSS}
\end{eqnarray}
all follow from Eqs.(\ref{NS1}), (\ref{NS2}) and (\ref{NS3}).

The average flux of left handed electron neutrinos (presumed massless) 
is determined by 
\begin{equation}
{\cal F}^\mu (x)=c\left<\bar{\nu }(x)\gamma^\mu P_+\nu (x)\right>.
\label{NS4}
\end{equation}  
Initial state averaging in Eq.(\ref{NS4}) is with respect to a chemical cell 
density matrix 
\begin{eqnarray}
\left<\ldots \right>\equiv Tr\ \rho \left(\ldots \right),
\nonumber \\ 
\rho =\sum_I p_I \left| I \right>\left< I \right|.
\label{NS4density}
\end{eqnarray}
The mean number of neutrinos created per unit time per unit volume may be 
computed from the four divergence of the neutrino flux; i.e.  
\begin{equation}
\varpi(x) =\partial_\mu  {\cal F}^\mu (x).
\label{NS5}
\end{equation}
Let us now argue, purely from standard model conservation laws, that 
\begin{math} \varpi  \end{math} is also the mean number of 
neutrons created per unit time per unit volume within the metallic hydride 
cathode in a chemical cell. 

If a neutrino is created, then {\em lepton number conservation} dictates that 
an electron had to be destroyed. If an electron is destroyed, then 
{\em charge conservation} dictates that a proton had to be destroyed. 
If a proton is destroyed, then {\em baryon number conservation} dictates that a neutron 
had to be created. Thus, the rate of neutrino creation must be equal to the rate 
of neutron creation. It is theoretically simpler to keep track 
of neutrino creation within the cathode.

The neutrino sinks and sources, respectively \begin{math} \bar{\eta}  \end{math} 
and \begin{math} \eta  \end{math}, are defined by that part of the standard model 
action which destroy and create neutrinos; i.e. 
\begin{equation}
S_{\rm int}=\hbar \int \left(\bar{\eta }(x)\nu (x)
+\bar{\nu}(x)\eta (x)\right)d^4x.
\label{NS6}
\end{equation} 
The neutrino field equations are thereby 
\begin{eqnarray}
-i\gamma^\mu \partial_\mu \nu (x)=\eta (x),
\nonumber \\ 
i \partial_\mu \bar{\nu }(x)\gamma^\mu =\bar{\eta }(x). 
\label{NS7}
\end{eqnarray} 
Eqs.(\ref{NS4}), (\ref{NS5}) and (\ref{NS7}) imply the neutrino 
creation rate per unit time per unit volume at space-time point 
\begin{math} x \end{math} has the form  
\begin{equation}
\varpi (x)=2c{\Im m}\left<\bar{\eta }(x)P_+\nu (x)\right>.
\label{NS8}
\end{equation}
Introducing the retarded massless Dirac Green's function,  
\begin{equation}
-i\gamma^\mu \partial_\mu S(x-y)=\delta (x-y),
\label{NS9}
\end{equation} 
allows us to solve the neutrino field Eqs.(\ref{NS7}) in the form 
\begin{equation}
\nu (x)=\nu_{in}(x)+\int S(x-y)\eta (y)d^4y,
\label{NS10}
\end{equation} 
wherein \begin{math} \nu_{in}(x) \end{math} represents the asymptotic 
incoming neutrino field. The assumption of {\em zero initial 
background neutrinos} is equivalent to the mathematical statement that  
the neutrino destruction operator   
\begin{math} \nu_{in}^+(x)\left|I\right>=0 \end{math} for the initial  
states in Eq.(\ref{NS4density}). 
In such a case, Eqs.(\ref{NS8}) and (\ref{NS10}) imply
\begin{equation} 
\varpi (x)=2c{\Im m}\int \left<\bar{\eta }(x)P_+S(x-y)\eta (y)\right>d^4y.
\label{NS11}
\end{equation}

The retarded massless Dirac Green's function may be found by looking 
for a solution of Eq.(\ref{NS9}) of the form 
\begin{equation}
S(x-y)=i\gamma^\mu \partial_\mu \Delta (x-y). 
\label{NS12}
\end{equation}
From Eqs.(\ref{NS9}) and (\ref{NS12}) it follows that 
\begin{equation}
-\partial_\mu \partial^\mu \Delta (x-y)=\delta (x-y).
\label{NS13}
\end{equation}
The retarded solution to Eq.(\ref{NS13}) requires the step function 
\begin{eqnarray}
\vartheta (x-y)=1\ \ {\rm if}\ \ x^0 > y^0,
\nonumber \\ 
\vartheta (x-y)=0\ \ {\rm if}\ \ x^0 < y^0;
\label{NS14}
\end{eqnarray}
In detail 
\begin{equation}
\Delta (x-y)=\frac{\vartheta (x-y)}{2\pi}
\ \delta \left((x-y)^2\right).
\label{NS15}
\end{equation}
Eqs.(\ref{NS11}) and (\ref{NS12}) imply 
\begin{eqnarray}
\varpi (x)=2c{\Re e}\int 
\left<\bar{\eta }(x)P_+\gamma^\mu \eta (y)\right>
\partial_\mu \Delta (x-y)d^4y,
\nonumber \\ 
\varpi (x)=2c{\Re e}\int \Delta (x-y)
 \left<\bar{\eta }(x)P_+\gamma^\mu \partial_\mu \eta (y)\right>d^4y.
\label{NS16}
\end{eqnarray}
The neutron production rate 
\begin{math} \varpi  \end{math} per unit time per unit volume 
can thus be computed in terms of the 
neutrino sinks \begin{math} \bar{\eta}  \end{math} and 
sources \begin{math} \eta  \end{math}.

\section{Composite Charged Fields \label{CCF}}

The neutrino sinks and sources of interest in this work 
can be written in terms of the composite fields of charged 
electrons and charged effective condensed matter 
\begin{math} W^\pm \end{math}-bosons; i.e. 
\begin{eqnarray}
\eta (y)=\frac{1}{\sqrt{2}}\gamma^\sigma W^+_\sigma (y) 
P_+\psi (y)\ ,
\nonumber \\ 
\bar{\eta }(x)=\frac{1}{\sqrt{2}}\bar{\psi }(x)
P_-\gamma^\lambda W^-_\lambda (x)\ ,
\label{CCF1}
\end{eqnarray}
in which \begin{math} \psi \end{math} and 
\begin{math} \bar{\psi}\end{math} are the Dirac electron fields 
and 
\begin{eqnarray}
W^+_\sigma (y)= \left(\frac{2\hbar G_F}{c^4}\right){\cal I}^+_\sigma (y)\ ,
\nonumber \\ 
W^-_\lambda (x)=\left(\frac{2\hbar G_F}{c^4}\right){\cal I}^-_\lambda (x)  \ .
\label{CCF1W}
\end{eqnarray}

The weak interaction proton-neutron charged conversion currents 
\begin{math} {\cal I}^\pm_\mu  \end{math} are defined in Eq.(\ref{intro7}). 
In Feynman diagram language, the amplitude pictured in FIG.\ref{fig1} has been 
replaced via a field current identity of the Fermi four field point 
interaction in FIG.\ref{fig2}.
Eqs.(\ref{NS16}) and (\ref{CCF1}) imply
\begin{equation}
\varpi (x)=c\ {\Re e}\int 
{\cal G}(x,y) \Delta (x-y)d^4y, 
\label{CCF2}
\end{equation}
wherein 
\begin{eqnarray}
{\cal G}(x,y) =
\ \ \ \ \ \ \ \ \ \ \ \ \ \ \ \ \ \ \ \ 
\nonumber \\ 
\left<W^-_\lambda (x)\bar{\psi }(x)\gamma^\lambda 
\gamma^\mu \gamma^\sigma P_+\partial_\mu \big(\psi (y)W^+_\sigma (y)\big)\right>=
\nonumber \\ 
h^{\lambda \mu \sigma \nu }\left<W^-_\lambda (x)\bar{\psi }(x)
\gamma_\nu \partial_\mu \big(P_+\psi (y)W^+_\sigma (y)\big)\right>.
\label{CCF3}
\end{eqnarray}
The neutron production rate \begin{math} \varpi \end{math} per unit time per 
unit volume implicit in Eqs.(\ref{NS15}), (\ref{CCF2}) and (\ref{CCF3}) may be 
considered to be exact. 

\section{Electron Mass Renormalization \label{EMR}}

\begin{figure}[tp]
\scalebox {0.6}{\includegraphics{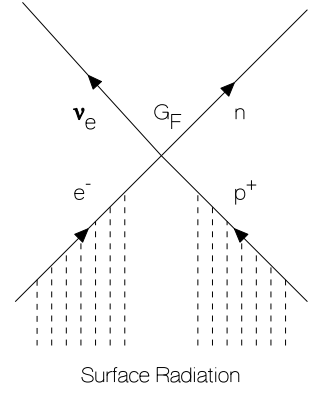}}
\caption{In the presence of electromagnetic surface radiation, the charged 
particles in weak interaction must be described by wave functions which 
include to high order in $\alpha $ the effects of the electromagnetic 
fields. For the reaction at hand, both the proton and the electron react 
to surface radiation. The resulting mass renormalization is stronger for 
the electronic degrees of freedom than for the proton degrees of freedom. 
The density of states including radiation is computed employing 
Eqs.(\ref{EMR4}) and (\ref{EMR5}).}
\label{fig3}
\end{figure}

When the reacting charged particles  
\begin{math} e^- +p^+ \to n+\nu_e \end{math} 
are in the presence of surface plasmon radiation, then the external 
charged lines (incoming wave functions) must include the radiation 
fields to a high order in the quantum electrodynamic coupling 
strength\cite{Lifshitz:1997} \begin{math} \alpha  \end{math}. 
The situation is shown in FIG.\ref{fig3}.
To see what is involved, recall how one calculates the density of states 
for two particles incoming and two particles outgoing:
\par \noindent  
{\it Case I: The Vacuum} The density of states may be written as the four 
momentum conservation law, 
\begin{eqnarray}
\int e^{i(S_{0+}+S_{0-}-S_{0n}-S_{0\nu})/\hbar}d^4x= 
\nonumber \\ 
(2\pi \hbar )^4\delta (p_++p_--p_n-p_\nu ),
\label{EMR1}
\end{eqnarray} 
wherein \begin{math} p_+ \end{math}, \begin{math} p_- \end{math}, 
\begin{math} p_n \end{math} and \begin{math} p_\nu \end{math} represent,  
respectively, the four momenta of the proton, electron, neutron and neutrino. 
For a particle of mass \begin{math} m \end{math} in the vacuum, the action 
\begin{math} S_0(x) \end{math} obeys the Hamilton-Jacobi equation 
\begin{eqnarray}
\partial_\mu S_0(x) \partial^\mu S_0(x)+m^2c^2=0, 
\nonumber \\ 
S_0(x)=p\cdot x\equiv p_\mu x^\mu .
\label{EMR2}
\end{eqnarray}  
{\it Case II: Radiation} If the reaction takes place 
in the presence of electromagnetic radiation, 
\begin{equation}
F_{\mu \nu }=\partial_\mu A_\nu - \partial_\nu A_\mu ,
\label{EMR3}
\end{equation}
then the density of states conservation of four momenta must 
also include the electromagnetic radiation contribution; i.e.  
\begin{eqnarray}
{\Re e}\int e^{i(S_{+}+S_{-}-S_{n}-S_{\nu})/\hbar}d^4x= 
\nonumber \\ 
(2\pi \hbar )^4\tilde{\delta }(p_+,p_-,p_n,p_\nu ),
\label{EMR4}
\end{eqnarray} 
wherein, for a charged particle, the Hamilton-Jacobi equation
reads\cite{Landau:1975} 
\begin{eqnarray}
mv_\mu (x)=\partial_\mu S(x)-\frac{e}{c}A_\mu (x), 
\nonumber \\  
v_\mu (x)v^\mu (x)+c^2=0.
\label{EMR5} 
\end{eqnarray}
Therefore, in the density of states Eq.(\ref{EMR4}) 
including radiation the full solution of the Hamilton-Jacobi 
equation must be solved for all of the charged particles in the 
interaction. This constitutes the physical difference between the 
diagrams in the vacuum shown in FIG.\ref{fig2} and including radiation 
shown in FIG.\ref{fig3}. Under a gauge transformation 
\begin{math} A_\mu \to A_\mu  + \partial_\mu \chi \end{math}, 
the Hamilton-Jacobi Eq.(\ref{EMR5}) for a charged particle 
implies a change in the action 
\begin{math} S\to S+e\chi /c  \end{math}.
The renormalized charged particle wave function   
thereby exhibits the expected gauge transformation rule 
\begin{math} \psi \to \psi \exp(e\chi /\hbar c ) \end{math} 
making the complete amplitude from FIG.\ref{fig3} gauge invariant. 
The electron gauge contribution to the phase is canceled 
by the proton gauge contribution to the phase since the two particles 
are oppositely charged.  

The mass renormalization may be understood 
by averaging the local momentum 
\begin{math} p_\mu = \partial_\mu S \end{math} 
over local space time regions. Presuming 
\begin{math} \overline{p^\mu A_\mu} =0 \end{math} 
we have on average that  
\begin{equation}
-\overline{p_\mu p^\mu}=m^2c^2+
\left(\frac{e}{c}\right)^2 \overline{A_\mu A^\mu}
\equiv \tilde{m}^2c^2.
\label{EMR5a}
\end{equation}
The mass renormalization parameter in Eq.(\ref{est1}) is then 
given by 
\begin{equation}
\beta=\sqrt{1+\left(\frac{e}{mc^2}\right)^2 \overline{A_\mu A^\mu}}\ .
\label{EMR5b}
\end{equation}
Since the electron mass is much less than the proton 
mass, \begin{math} m\ll M_p  \end{math}, the main effects 
on low energy nuclear reactions are due to 
the mass renormalization of the surface electrons\cite{Widom:2006}. 

From the viewpoint of classical physics, the Lorentz force on a charge 
equation of motion, 
\begin{eqnarray}
mc\frac{dv^\mu }{d\tau } = eF^{\mu \nu }v_\nu ,
\nonumber \\ 
v^\mu = \left(\frac{\bf v}{\sqrt{1-(v/c)^2}},\frac{c}{\sqrt{1-(v/c)^2}}\right),
\label{EMR6}
\end{eqnarray}
is reduced to first order via the Hamilton-Jacobi Eq.(\ref{EMR5}), 
according to
\begin{equation}
\frac{dx^\mu }{d\tau }=v^\mu (x).
\label{EMR7}
\end{equation}
From the viewpoint of quantum mechanics, there is a one to one correspondence 
between quantum solutions of the Dirac equation and the classical solutions of 
Hamilton-Jacobi equation. In detail, the Dirac equation in an external radiation 
field,
\begin{equation}
-i\hbar \gamma^\mu 
\left\{\partial_\mu -i\left(\frac{e}{\hbar c}\right)A_\mu (x)\right\}\psi (x)
+mc\psi (x)=0,
\label{EMR8}
\end{equation} 
may be subject to a non-linear gauge transformation employing the solution to the 
classical Hamilton-Jacobi equation, 
\begin{equation}
\psi(x)= e^{iS(x)/\hbar }\Psi(x).
\label{EMR9}
\end{equation} 
The resulting radiation renormalized wave function obeys 
\begin{equation}
\gamma^\mu \left(-i\hbar \partial_\mu +mv_\mu (x)\right)\Psi (x)
+mc\Psi(x)=0.
\label{EMR10}
\end{equation}
It is worthy of note in the quasi-classical limit 
\begin{math} \hbar \to 0 \end{math} that the solution to 
the charged particle wave Eq.(\ref{EMR10}) is reduced to algebra. 

\section{Proton Oscillations \label{EPO}}

Thermal neutron scattering from  
hydrogen\cite{Dreismann:2005,Cowley:2006}, metallic 
hydrides\cite{Kenali:1999,Kolesnikova:2002,Glagolenkol:2002,Gidopoulos:2005}  
and even from protons embedded in proteins\cite{Medini:2003} has 
been of considerable recent theoretical and 
experimental interest\cite{Platzman:2004,Krzystyniak:2006}.
Employing the total cross section per unit volume as the extinction coefficient, 
\begin{equation}
h_{tot}=\frac{\sigma_{tot}}{\cal V},
\label{EPO1}
\end{equation} 
the intensity of a neutron beam when passed through a sample of thickness 
\begin{math} d \end{math} obeys 
\begin{equation}
I(d)=I(0)\exp(-h_{tot}d).
\label{EPO2}
\end{equation} 
The neutron-proton cross section is quite large since there is a very shallow bound 
state (the deuteron). For high densities of protons, the neutron-proton scattering 
will normally dominate the condensed matter scattering of thermal neutrons.  
Let the momentum transfer and energy transfer to the neutron 
be defined, respectively, as 
\begin{equation}
\hbar {\bf Q}={\bf p}_i-{\bf p}_f\ \ \ {\rm and}
\ \ \ \hbar \omega=\epsilon_i-\epsilon_f.
\label{EPO3}
\end{equation}
Decomposing the neutron scattering into various momementa and energies yields 
\begin{equation}
\frac{d^2 h_{i\to f}}{d\Omega_f d\epsilon_f}=\frac{\cal{N}}{\hbar {\cal V}}\left(\frac{p_f}{p_i}\right)
\frac{d \sigma_{i\to f}}{d\Omega_f}S({\bf Q},\omega ),
\label{EPO4}
\end{equation}
wherein \begin{math} d \sigma_{i\to f} \end{math} is the differential cross section for 
a single neutron to scatter off a single proton and the dynamical proton form factor for 
\begin{math} {\cal N}  \end{math} protons is defined as 
\begin{eqnarray}
{\cal N}S({\bf Q},\omega )=\sum_{j=1}^{\cal N}\int_{-\infty }^\infty e^{i\omega t}
F_j({\bf Q},t)\frac{dt}{2\pi }
\nonumber \\ 
F_j({\bf Q},t)=\left<e^{-i{\bf Q\cdot R}_j(t)}e^{i{\bf Q\cdot R}_j(0)}\right>
\label{EPO5}
\end{eqnarray}
In what follows we shall consider the self diffusion for a fixed proton at position 
\begin{math} {\bf R}_j(t)\equiv  {\bf R}(t) \end{math} and write 
\begin{equation}
\left<e^{-i{\bf Q\cdot R}(t)}e^{i{\bf Q\cdot R}(0)}\right>=\int_{-\infty }^\infty 
S({\bf Q},\omega )e^{-i\omega t}d\omega .
\label{EPO6}
\end{equation}
Under the assumption that the proton motions are non-relativistic, one easily shows for 
the many body Hamitonian \begin{math} {\cal H} \end{math} that  
\begin{equation}
{\cal H}_{\bf Q}=e^{-i{\bf Q\cdot R}}{\cal H}e^{i{\bf Q\cdot R}}=
{\cal H}+\hbar {\bf V\cdot Q}+\frac{\hbar^2 Q^2}{2M},
\label{EQO7}
\end{equation} 
wherein the velocity of the proton is 
\begin{equation}
{\bf V}=-i\left(\frac{\hbar }{M}\right){\bf \nabla}.
\label{EPO8}
\end{equation}
Thus 
\begin{eqnarray}
F({\bf Q},t)&=&\left<e^{-i{\bf Q\cdot R}(t)}e^{i{\bf Q\cdot R}(0)}\right>,
\nonumber \\
F({\bf Q},t)&=&\left<e^{i{\cal H}t}e^{-i{\bf Q\cdot R}}
e^{-i{\cal H}t}e^{i{\bf Q\cdot R}}\right>,
\nonumber \\
F({\bf Q},t)&=&\left<e^{i{\cal H}t/\hbar}e^{-i{\cal H}_{\bf Q}t/\hbar}\right>,
\nonumber \\
F({\bf Q},t)&=&e^{-i\omega_Qt}\left<e^{-i\int_0^t {\bf Q\cdot V}(s)ds}\right>_+ ,
\label{EPO9}
\end{eqnarray}
wherein ``+'' indicates time ordering, 
\begin{equation}
\omega_{\bf Q}=\frac{\hbar Q^2}{2M}\ \ \ {\rm and}
\ \ \ {\bf V}(s)=e^{i{\cal H}s/\hbar }{\bf V}e^{-i{\cal H}s/\hbar }.
\label{EPO10}
\end{equation}
From Eqs.(\ref{EPO6}), (\ref{EPO9}) and (\ref{EPO10}) one derives the sum rules 
\begin{eqnarray}
\int_{-\infty}^\infty S({\bf Q},\omega )d\omega &=&1,
\nonumber \\
\int_{-\infty}^\infty \omega S({\bf Q},\omega )d\omega &=&\omega_{\bf Q},
\label{EPO11}
\end{eqnarray}
which represent, respectively, probability normalization and the fact that the mean 
recoil energy is the same as would have been computed for a single free proton.
We can now discuss the experimental measurements which make proton oscillations 
unusual. 

\subsection{Deep Inelastic Neutron Scattering} 

For energetic neutrons with relatively 
high momentum transfer, it was estimated that the neutron scattering event 
lasted on a time scale of an attosecond. It was thought that the velocity could not appreciably change in so short a time. Therefore one might try the {\em impulse} approximation wherein Eq.(\ref{EPO9}) could be estimated by the 
\begin{math} t\to 0  \end{math} limit   
\begin{eqnarray}
F_{\rm impulse}({\bf Q},t)=e^{-i\omega_Qt}\left<e^{-i {\bf Q\cdot V}t}\right>,
\nonumber \\
S_{\rm impulse}({\bf Q},\omega )=
\int \delta (\omega -{\bf Q\cdot V}-\omega_{\bf Q})p({\bf V})d^3{\bf V},
\label{EPO12}
\end{eqnarray}
wherein \begin{math} p({\bf V})d^3{\bf V} \end{math} is the probability that the proton 
velocity is in the range \begin{math} {\bf V}\in d^3{\bf V} \end{math}.
Eq.(\ref{EPO12}) is at the heart of the theoretical analysis of so-called ``deep inelastic'' 
neutron scattering. Note that the impulse approximation obeys the sum rules in 
Eq.(\ref{EPO11}). It came as somewhat of a shock that for protons inside liquid hydrogen, 
metallic hydrides and proteins, the impulse approximation does not work. At issue was the 
``violation'' of the probability sum rule
\begin{equation}
\int_{\rm experimental}S(Q,\omega )d\omega <1.
\label{EPO13}
\end{equation}
Since Eqs.(\ref{EPO1}) and (\ref{EPO2}) were experimentally verified, the only reason for 
the ``experimental probability loss'' must have been that the integral over experimental 
high frequency data was missing a physical low frequency regime which persists and 
contributes to the probability integral. Such low frequency modes must be collective.

\subsection{Recoilless Fraction} 

Let us at first assume the cluster decomposition property 
of \begin{math} F({\bf Q},t) \end{math}; i.e. 
\begin{eqnarray}
\lim_{t\to \infty}F({\bf Q},t)=F_{\infty}({\bf Q}),
\nonumber \\ 
\lim_{t\to \infty}\left<e^{-i{\bf Q\cdot R}(t)}e^{i{\bf Q\cdot R}(0)}\right>
=\left|\left<e^{i{\bf Q\cdot R}}\right>\right|^2, 
\nonumber \\ 
F_{\infty}({\bf Q})=\left|\left<e^{i{\bf Q\cdot R}}\right>\right|^2. 
\label{EPO14}
\end{eqnarray}
If cluster decomposition holds true, then \begin{math} F_{\infty }({\bf Q}) \end{math}
represents the {\em recoilless fraction} in the sense that 
\begin{eqnarray}
S({\bf Q},\omega )=F_{\infty}({\bf Q})\delta (\omega )+\tilde{S} ({\bf Q},\omega )
\nonumber \\ 
\int_{-\infty}^\infty \tilde{S}({\bf Q},\omega )d\omega =1-F_{\infty}({\bf Q})<1. 
\label{EPO15}
\end{eqnarray}
Since the second of Eqs.(\ref{EPO15}) yield the fraction that recoils, the experimental 
puzzle in Eq.(\ref{EPO13}) is then resolved. 

The recoilless fraction \begin{math}  F_{\infty}({\bf Q}) \end{math} for the proton 
is the same factor as the recoilless fraction \begin{math}  F_{\infty}({\bf Q}) \end{math} 
that appears in the M\"ossbauer effect for the gamma ray decay in a heavy nucleus.  
In the case of the M\"ossbauer effect, the recoil of the gamma emission is taken 
up by the crystal as a whole.  For the case of deep inelastic neutron scattering, 
the proton recoil must be taken up by neighboring electrons and other protons in 
coherent oscillation. Let \begin{math} P({\bf u})d^3 {\bf u} \end{math} be the 
probability that the displacement of a proton from its equilibrium position
\begin{math} {\bf u}={\bf R}-\big<{\bf R}\big>  \end{math} is in the range 
\begin{math} {\bf u}\in d^3 {\bf u} \end{math}. The recoilless fraction in 
Eq.(\ref{EPO14}) may be written as 
\begin{equation}
F_{\infty}({\bf Q})=
\left|\int P({\bf u})e^{i{\bf Q\cdot u}}d^3{\bf u}\right|^2.
\label{EPO16}
\end{equation}
The mean square fluctuations in displacement as defined by 
\begin{equation}
\overline{\bf uu}=\int P({\bf u}) {\bf uu} d^3{\bf u},
\label{EPO17}
\end{equation}  
can be measured via the low momentum transfer limit 
\begin{equation}
\lim_{Q\to 0}\frac{\overline{|{\bf Q\cdot u}|^2}}{Q^2}
=-\lim_{Q\to 0}\frac{\ln F_{\infty}({\bf Q})}{Q^2}.
\label{EPO18}
\end{equation}  
For higher momentum transfers, the fractal dimensional spectator 
model\cite{Medini:2003} gives a fairly good representation of the 
recoilless fraction; i.e.  
\begin{equation}
F_{\infty}({\bf Q};D)=
\left[1+\frac{\overline{|{\bf Q\cdot u}|^2}}{D}\right]^{-D}.
\label{EPO19}
\end{equation}
The harmonic oscillation result is the Gaussian probability Debye-Waller factor  
\begin{equation}
\lim_{D\to \infty }F_{\infty}({\bf Q};D)=
\exp\left(-\overline{|{\bf Q\cdot u}|^2}\right).
\label{EPO20}
\end{equation}
At elevated temperatures, a small value of \begin{math} D \end{math} describes the 
fractal dimension of the probability distribution \begin{math} P({\bf u})d^3 {\bf u} \end{math}.
Finally, for the case of collective proton oscillations, the delta function in 
Eq.(\ref{EPO15}) is somewhat idealized, In experimental practice, the delta function 
is broadened into an infrared peak consistant in thermal equilibrium with the detailed 
balance condition 
\begin{equation}
S(-{\bf Q},-\omega )=e^{-\hbar \omega /k_BT}S({\bf Q},\omega ).
\label{EPO21}
\end{equation}  

\subsection{Proton Displacements} 

Neutron scattering experiments\cite{Kenali:1999} on 
palladium hydride at moderate momentum transfer clearly 
indicate a sharply defined collective oscillation peak at 
\begin{math} (\hbar \Omega /e)\approx 60\ {\rm millivolt} \end{math}. 
Such a collective proton oscillation at an infrared frequency will resonate with 
electronic surface plasmon oscillations of the electrons leading to the local 
breakdown of the Born-Oppenheimer approximation and large collective proton oscillation 
amplitudes. Some theoretical insights into the nature of such oscillations can be obtained 
from sum rules. The mobility tensor of a single proton at complex frequency may be defined 
via the Kubo formula 
\begin{equation}
{\sf m}(\zeta )=\frac{i}{\hbar }\int_0^\infty 
\left<\left[{\bf V}(t),{\bf R}(0)\right]\right>e^{i\zeta t}dt,
\ \ \ {\Im m}(\zeta )>0. 
\label{EPO22}
\end{equation}
The mobility tensor obeys the dispersion relation 
\begin{equation}
{\sf m}(\zeta )=\left(\frac{-2i\zeta }{\pi }\right)
\int_0^\infty \frac{\Re e{\sf m}(\omega +i0^+)d\omega }{\omega^2-\zeta^2}.
\label{EPO23}
\end{equation}
As the complex frequency \begin{math} |\zeta |\to \infty  \end{math} in the 
upper half plane, the mobility tensor obeys the asymptotic form 
\begin{eqnarray}
{\sf m}(\zeta )=
\left(\frac{2i}{\pi \zeta }\right)\int_0^\infty \Re e{\sf m}(\omega +i0^+)d\omega +
\nonumber \\ 
\left(\frac{2i}{\pi \zeta^3 }\right)\int_0^\infty 
\omega^2 \Re e{\sf m}(\omega +i0^+)d\omega +\ldots 
\label{EPO24}
\end{eqnarray}
The same asymptotic expansion can be obtained by continually integrating 
Eq.(\ref{EPO22}) by parts yielding the sum rules 
\begin{eqnarray}
\frac{2}{\pi }\int_0^\infty \Re e{\sf m}(\omega +i0^+)d\omega =
\frac{i}{\hbar }\left<\left[{\bf V},{\bf R}\right]\right>,
\nonumber \\ 
\frac{2}{\pi }\int_0^\infty \omega^2 \Re e{\sf m}(\omega +i0^+)d\omega =
\frac{i}{\hbar }\left<\left[{\bf A},{\bf V}\right]\right>,
\label{EPO25}
\end{eqnarray}
wherein the proton velocity is \begin{math} {\bf V}=\dot{\bf R}  \end{math} 
and the proton acceleration is \begin{math} {\bf A}=\dot{\bf V}  \end{math}.
Employing the {\em non-relativistic proton} Eqs.(\ref{EPO8}), (\ref{EPO25}) and 
Newton's law \begin{math} M{\bf A}=|e|{\bf E} \end{math}, we find that
\begin{eqnarray}
\frac{2}{\pi }\int_0^\infty \Re e{\sf m}(\omega +i0^+)d\omega =
\frac{\sf 1}{M},
\nonumber \\ 
\frac{2}{\pi }\int_0^\infty \omega^2 \Re e{\sf m}(\omega +i0^+)d\omega =
-\frac{|e|}{M^2}\left<{\bf \nabla E}\right>.
\label{EPO26}
\end{eqnarray}
Employing Gauss'law at the proton position 
\begin{math} \big<div{\bf E}\big>=-|e|\tilde{n} \end{math} 
wherein \begin{math} \tilde{n} \end{math} is the electron density right on top 
of the proton, 
\begin{equation}
\tilde{n}=\left<\sum_k \delta ({\bf R}-{\bf r}_k)\right>
=\left<\psi^\dagger ({\bf R})\psi ({\bf R})\right>,
\label{EPO27}
\end{equation}
the sum rule estimate for the infrared frequency peak in the neutron scattering is 
thereby
\begin{eqnarray}
\Omega^2=\frac{\int_0^\infty \omega^2 \Re e\ tr\ {\sf m}(\omega +i0^+)d\omega }
{\int_0^\infty \Re e\ tr\ {\sf m}(\omega +i0^+)d\omega }\ ,
\nonumber \\ 
\Omega^2=\frac{e^2\tilde{n}}{3M}=
\frac{4\pi }{3}\left(\frac{\hbar^2}{Mm}\right)\frac{\tilde{n}}{a}\ .
\label{EPO28}
\end{eqnarray}
In Eq.(\ref{EPO28}), \begin{math} a \end{math} is the Bohr radius. 
In order to understand more clearly our estimate of the proton oscillation 
frequency \begin{math} \Omega  \end{math}, suppose that the proton was embedded 
in a sphere with charge density \begin{math} -|e|\tilde{n}  \end{math}. If the 
proton is pulled out by a displacement \begin{math} {\bf u} \end{math}, then an 
electric field would exist  
\begin{equation}
|e|{\bf E}=-\left(\frac{e^2\tilde{n}}{3}\right){\bf u}=-M\Omega^2{\bf u}
\label{EPO29}
\end{equation}
pushing the proton back to the sphere center. The equation of motion 
\begin{math}M\ddot{\bf u}=-|e|{\bf E}=-M\Omega^2{\bf u} \end{math} yields 
the required oscillation. One may also note the identity 
\begin{equation}
e^2\overline{|{\bf E}|^2}=M^2\Omega^4\overline{|{\bf u}|^2}
\label{EPO30}
\end{equation}
relating the meansquare electric field at the proton position to 
the mean field displacement fluctuation. Suppose a momentum for a 
relativistic electron given by 
\begin{math} {\bf p}=m{\bf v}/\sqrt{1-|{\bf v}/c|^2}  \end{math}.
The energy \begin{math} K \end{math} of such an electron may 
then be written 
\begin{equation}
\overline{K^2}=m^2c^4+\overline{|{\bf p}|^2}c^2
\label{EPO31}
\end{equation}
equivalent to 
\begin{equation}
\overline{K^2}=m^2c^4+\frac{e^2\overline{|{\bf E}|^2}c^2}{\Omega^2} 
=m^2c^4\left[1+\frac{\overline{|{\bf E}|^2}}{{\cal E}^2}\right].
\label{EPO32}
\end{equation}
wherein the electric field scale has been introduced 
\begin{equation}
{\cal E}=\frac{mc}{\hbar}\left(\frac{\hbar \Omega}{e}\right).
\label{EPO32a}
\end{equation}

\begin{figure}[tp]
\scalebox {0.7}{\includegraphics{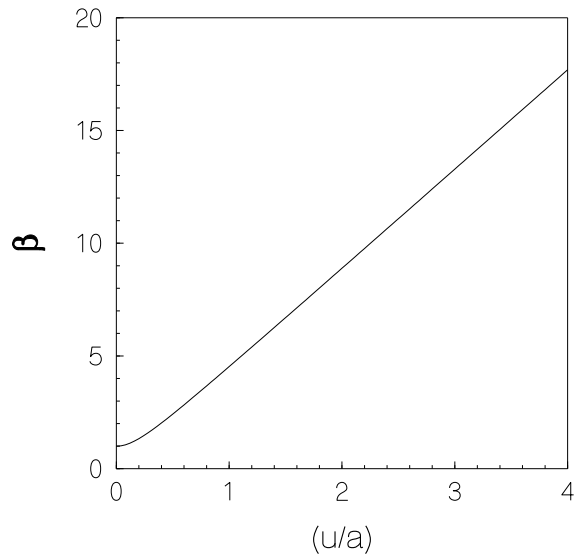}}
\caption{The predicted electron mass enhacemnt $\beta =\tilde{m}/m$ when 
protons are absorbed into palladium is plotted as a function of the root 
mean square proton displacement $u$ where $u^2=\overline{|{\bf u}|^2}$ and  
the Bohr radius $a\approx 0.5291772108\times 10^{-8}$ cm.}
\label{fig4}
\end{figure}

Experimentally\cite{Kenali:1999}, for protons absorbed in Pd, 
one finds the soft Boson mode with  
\begin{math} (\hbar \Omega /e)\approx 6\times 10^{-2}\ {\rm volts} \end{math} 
wherein 
\begin{equation}
{\cal E}_{\rm p \in Pd}\approx 1.55\times 10^{11}\ {\rm volts/meter}.
\label{EQO32b}
\end{equation}
In this regard, one notes that typical atomic physics electric fields in 
atoms are of that order when located a Bohr radius 
\begin{math} a \end{math} from an isolated proton; i.e. with the Bohr 
radius  
\begin{equation}
a=\frac{4\pi \hbar^2}{e^2m}=
\frac{\hbar }{\alpha mc}\approx 0.5291772108\times 10^{-8}\ {\rm cm},
\label{EQO36}
\end{equation}
atomic electric fields are of the order 
\begin{equation}
{\cal E}_a = \frac{|e|}{4\pi a^2}\approx 
5.142206318\times 10^{11} \ {\rm volts/meter}.
\label{EQO32c}
\end{equation}
Employing Eqs.(\ref{EPO28}), (\ref{EPO30}) and (\ref{EQO32c})  
along with the estimate 
\begin{math} \tilde{n}\approx |\psi(0)|^2=1/(\pi a^3) \end{math}  
yields  
\begin{equation}
\frac{\overline{|{\bf E}|^2}}{{\cal E}_a^2}=\frac{16}{9}
\left(\frac{\overline{|{\bf u}|^2}}{a^2}\right).
\label{EQO33}
\end{equation}
Thus, Eq.(\ref{EMR5b}) may be written\cite{Widom:2006} as 
\begin{eqnarray}
\beta=\sqrt{\frac{\overline{K^2}}{m^2c^4}}
=\sqrt{1+\left[\frac{e^2\overline{|{\bf E}|^2}}{m^2c^2\Omega^2}\right]}\ ,
\nonumber \\ 
\beta =\frac{\tilde{m}}{m}\approx
\sqrt{1+\frac{16{\cal E}_a^2}{{9{\cal E}}_{\rm p}^2}
\left(\frac{\overline{|{\bf u}|^2}}{a^2}\right)}\ .
\label{EPO33}
\end{eqnarray}
For example, for protons in palladium Eqs.(\ref{EQO32b}) and (\ref{EQO32c}) 
imply
\begin{equation}
\beta_{\rm p \in Pd}\approx \sqrt{1+19.6
\left(\frac{\overline{|{\bf u}|^2}}{a^2}\right)},
\label{EPO34}
\end{equation}
which is shown in FIG.\ref{fig4}. 

\subsection{Driven Oscillations}

When an electrode has a current density passing into the cathode surface, then there 
is a power per unit area \begin{math} {\cal P}  \end{math} suppled to the surface 
as in Eq.(\ref{intro3}). Under such circumstance the surface temperature rises to a 
high value. In detail, in a non-equilibrium steady state situation, one must introduce 
a ``noise temperature'' \begin{math} T_{{\bf Q}, \omega}  \end{math} which depends on the 
frequency and wave number of the excitations which form due to an Ohmic heating rate 
per unit time per unit area \begin{math} {\cal P}={\cal V}{\cal J} \end{math}. The noise 
temperature enters into the dynamic form factor via 
\begin{equation}
S(-{\bf Q},-\omega )=e^{-\hbar \omega /k_BT_{{\bf Q}, \omega}}S({\bf Q},\omega )
\label{EPO35}
\end{equation}
which is the non-equilibrium noise temperature version of the thermal equilibrium detailed 
balance Eq.(\ref{EPO21}).

Experimentally, the noise temperature is high in the infrared  
\begin{math} \omega  \end{math}-regime in a small surface domain size 
\begin{math} d \end{math} related to the wave number via 
\begin{math} d\sim Q^{-1} \end{math}. Note that the overall temperature over the whole 
electromagnetic bandwidth need not be very high, but in the infrared regime at 
\begin{math} \omega \sim \Omega  \end{math} there should be a sharp peak in the noise 
temperature. What has been observed\cite{Szpak:2003} for deuterons absorbed in palladium, 
are surface hot spots flashing in different domains analogous to the flashing of fire-flies 
during a dark evening in the country. The flashing turns on and off in apparently random 
spatial and temporal patterns. The hot spots occur when and where regions of the rough 
cathode surface form electromagnetic cavities of size and shape as to naturedly support the 
infrared radiation. These occur randomly on rough cathode surfaces.

During these flash events the nuclear transmutations can 
take place. The hot spots at any given time take up only a small fraction of the 
cathode surface. Increasing the relative fraction of nuclear-active cathode surface 
area which comprises hot spots would proportionately increase the measured efficiency 
of neutron production on the cathode.  The higher the average density of hot spots on 
a given cathode surface over time, the greater the amount of excess heat that would likely 
be observed at the cathode device level using gross thermal measurement techniques such as 
calorimetry.  From the nature of the surface damage on cathodes due to micron-scale hot 
spots, it is evident that the metal actually melts locally during a hot flash producing 
nuclear transmutations. Distinctive areas of obvious melting and explosive blow-out 
cratering features that are consistent with the presence of such hot spots are very commonly 
seen in post-experiment SEM images of the surfaces of various metallic cathode materials that 
include gold (melting point $1554.9\ ^oC$; boiling point $2963\ ^oC$); 
palladium (melting point $1554.9\ ^oC$; 
boiling point $2963\ ^oC$)\cite{Toriyabe:2006,Szpak:2005}; 
titanium (melting point $1064.18\ ^oC$; boiling point $2856\ ^oC$)\cite{Savvatimova:2006}, 
and tungsten (melting point $3422\ ^oC$; boiling point $5555\ ^oC$)\cite{Cirillo:2006}. 
Such evidence is consistent with the noise temperature at a typical value of 
the peak \begin{math} T_{{\bf Q},\omega }\sim 5\times 10^3\ ^oK \end{math}
corresponding to a proton displacement of \begin{math}(u/a)\sim 10 \end{math} within a domain. 
In accordance with FIG.\ref{fig4}, the mass enhancement is more than sufficent for 
the neutron production Eqs.(\ref{intro1}) and (\ref{intro2}).

\section{Neutrons and Isotopic Spin Waves\label{ISPW}}

The sources of the neutrinos are inhomogeneous in spatial regions near 
the surfaces of cathodes. Also, the neutrinos are so weakly interacting 
that after emission they are virtually unaware of the condensed matter. 
The neutrino on energy shell phase space 
\begin{math} Q^\nu =({\bf Q},|{\bf Q}|) \end{math}
has the Lorentz invariant phase space  
\begin{equation}
dL_{\bf Q}=\left[\frac{d^3 {\bf Q}}{2(2\pi )^3|{\bf Q}|}\right].
\label{ISPW1}
\end{equation}
Writing the neutrino emission part of Eqs.(\ref{CCF2}) and (\ref{CCF3}) 
as the phase space integral 
\begin{eqnarray}
\varpi (x)=-c\ {\Im m}\int \int e^{iQ\cdot (x-y)} 
h^{\lambda \mu \sigma \nu } Q_\mu \times
\nonumber \\ 
\left<W^-_\lambda (x)\bar{\psi }(x)
\gamma_\nu  P_+\psi (y)W^+_\sigma (y)\right>d^4 y dL_{\bf Q}
\nonumber \\ 
=-4c\left(\frac{\hbar G_F}{c^4}\right)^2\ {\Im m}
\int \int e^{iQ\cdot (x-y)} 
h^{\lambda \mu \sigma \nu } Q_\mu \times
\nonumber \\ 
\left<{\cal I}^-_\lambda (x)\bar{\psi }(x)P_-\gamma_\nu  
\psi (y){\cal I}^+_\sigma (y)\right>d^4 y dL_{\bf Q}\ .
\label{ISPW2}
\end{eqnarray}
Under the {\em assumption} that the initial proton spins are 
{\em uncorrelated} and that the free neutron density is dilute, 
considerable simplification can be made in estimating the rather daunting 
but rigorous Eq.(\ref{ISPW2}). The estimate for the inhomogeneous ultra 
low momentum neutron production rate per unit volume is 
\begin{eqnarray}
\varpi (x)\approx 
\left(\frac{\hbar G_F}{c^3}\right)^2
\left(\frac{2mc^2}{\hbar }\right)(g_V^2+3g_A^2)\times 
\nonumber \\   
\ {\Re e}\int \int e^{iQ\cdot (x-y)}
\left<{\cal T}^+ (x)\bar{\psi }(x)  
\psi (y){\cal T}^-(y)\right>d^4 y dL_{\bf Q},\ 
\label{ISPW3}
\end{eqnarray}
wherein Eq.(\ref{intro13}) has been invoked. 

If the neutrons are dilute, then it is sufficient to consider the creation 
of a single neutron from a proton, i.e. the propagation of the 
\begin{math} W^\pm  \end{math} within condensed matter. What is left of the heavy 
\begin{math} W^\pm  \end{math} boson is merely an isotopic spin wave.  
There is a superposition of amplitudes summed over all the possible protons 
within a patch which may be converted into a neutron. The isotopic spin wave creation and 
annihilation operators in the surface patch obey 
\begin{equation}
{\cal T}^\pm (x)\approx {\cal T}^\pm ({\bf x})e^{\mp i(c\Delta M)x^0/\hbar }
\label{ISPW4}
\end{equation}
with the neutron-proton mass difference determining the threshold value of the 
electron mass \begin{math} m \end{math} renormalization parameter 
\begin{math} \beta \end{math} i.e.  
\begin{equation}
\Delta M=M_n-M=\beta_0 m.
\label{ISPW5}
\end{equation}
For the creation of a single ultra low momentum neutron from non-relativistic protons
\begin{eqnarray}
\left<{\cal T}^+ (x)\bar{\psi }(x) \psi (y){\cal T}^-(y)\right> 
\Rightarrow 
\nonumber \\ 
\delta ({\bf x}-{\bf y})e^{-i(c\Delta M)(x^0-y^0)/\hbar }\times  
\nonumber \\ 
\left<p^\dagger (x)\bar{\psi }(x) \psi (y)p(y)\right>.
\label{ISPW6}
\end{eqnarray}
For steady state production rates, 
Eq.(\ref{ISPW3}) reads  
\begin{eqnarray} 
\frac{\hbar \varpi ({\bf r})}{mc^2}\approx 
2 \left(\frac{\hbar G_F}{c^3}\right)^2(g_V^2+3g_A^2)\times 
\nonumber \\
\ {\Re e}\int \int e^{i(c^2\Delta M +\hbar c|{\bf Q}|)t/\hbar}\times 
\nonumber \\ 
\left<p^\dagger ({\bf r})\bar{\psi }({\bf r}) 
\psi ({\bf r,t})p({\bf r,t})\right>(cdt) dL_{\bf Q}. 
\label{ISPW7}
\end{eqnarray}
Explicitly exhibiting the neutrino energy being radiated away 
in Eq.(\ref{intro2}), yields 
\begin{eqnarray}
\frac{\hbar \varpi ({\bf r})}{mc^2}\approx 
\frac{1}{2\pi ^2 c^2} \left(\frac{\hbar G_F}{c^3}\right)^2
(g_V^2+3g_A^2)\times 
\nonumber \\
\ {\Re e}\int_0^\infty  \int_{-\infty}^\infty 
e^{i(c^2\Delta M +\hbar c|{\bf Q}|)t/\hbar}\times 
\nonumber \\ 
\left<p^\dagger ({\bf r})\bar{\psi }({\bf r}) 
\psi ({\bf r,t})p({\bf r,t})\right>(cdt)(\omega d\omega ). 
\label{ISPW8}
\end{eqnarray}

The remaining correlation function in Eq.(\ref{ISPW7}) describes 
how an electron which finds itself directly on top of a proton 
propagates in time. The integral over time may be written 
\begin{eqnarray}
{\Re e}\int_{-\infty}^\infty \left<p^\dagger ({\bf r})\bar{\psi }({\bf r}) 
\psi ({\bf r,t})p({\bf r,t})\right> e^{iEt/\hbar }dt
\nonumber \\ 
=2\pi \hbar {\cal N}({\bf r})n_e({\bf r},E),
\label{ISPW9}
\end{eqnarray}
wherein \begin{math} {\cal N}({\bf r})  \end{math} is the mean 
density per unit volume of protons and 
\begin{math} n_e({\bf r},E) \end{math} is the mean collective  
density per unit volume per unit energy of electrons which sit 
directly on the protons. 

The steady state inhomogeneous production of 
neutrons per unit time per unit volume 
\begin{math} \varpi({\bf r})  \end{math}
as estimated in Eq.(\ref{ISPW8}); i.e. exhibiting the radiated 
neutrino energy \begin{math} \hbar \omega \end{math}, 
\begin{eqnarray}
\varpi({\bf r})\approx \frac{mc^2}{\pi \hbar}
\left(\frac{\hbar G_F}{c^3}\right)^2(g_V^2+3g_A^2)
{\cal N}({\bf r}){\cal K}({\bf r}),
\nonumber \\ 
{\cal K}({\bf r})=\frac{\hbar }{c}\int_0^\infty 
n_e({\bf r},E=c^2\Delta M+\hbar \omega) \omega d\omega ,
\label{ISPW10}
\end{eqnarray}
wherein \begin{math} n_e({\bf r},E)  \end{math} must be 
calculated including the surface radiation energy and the driving 
current through the cathode. 

The calculation of \begin{math} {\cal K} \end{math} depends on 
the detailed physical properties of the cathode surface as well 
as the flux \begin{math} \tilde{\Phi }  \end{math} per unit area 
per unit time of electrons determining the chemical cell current 
as in the power Eq.(\ref{intro3}). In the most simple model, 
let us consider a smooth surface with material properties and electron 
currents determining the neutron creation rate via the mass 
renormalization parameter \begin{math} \beta  \end{math} as defined 
in Eq.(\ref{EMR5b}). We note in passing that a smooth surface is not 
likely to be the best surface for producing neutrons since rough surfaces 
have patches wherein the surface plasma electromagnetic 
field oscillations will be very much more intense. However, the following 
smooth surface model will be employed for estimating the proper low energy 
nuclear reaction rates produced by electroweak interactions.   

To compute the density of surface electron states per unit area per 
unit energy, one may begin with a simple renormalized electron mass 
\begin{math} \tilde{m} \end{math} model and take 
\begin{equation}
g_2(E )=2\int 
\delta \left(E -\sqrt{c^2p^2+(\tilde{m}c^2)^2}\right)
\frac{d^2{\bf p}}{(2\pi \hbar)^2}\ ,
\label{ISPW11}
\end{equation}
i.e.   
\begin{equation}
g_2(E)=\frac{E}{\pi \hbar^2 c^2} \ .
\label{ISPW12}
\end{equation}
If such surface electron states are confined to a wave function 
width  \begin{math} l  \end{math} normal to the surface, then we 
have within the surface region, and after integrating over the 
emitted neutrino energy spectrum
\begin{eqnarray}
{\cal K}\approx \frac{\hbar }{lc} \int 
g_2(E=\beta_0 mc^2+\hbar \omega )\omega d\omega ,
\nonumber \\ 
{\cal K}\approx \frac{1}{2\pi l}\left(\frac{mc}{\hbar }\right)^3
(\beta -\beta_0)^2.
\label{ISPW13}
\end{eqnarray}

For a smooth surface, integrating the neutrino production rate over 
a thin slab at the electrode surface yields the estimate for 
the production rate per unit time per unit area, 
\begin{equation}
\varpi_2 \approx \left(\frac{g_V^2+3g_A^2}{2\pi^2}\right)
n_2\left(\frac{G_Fm^2}{\hbar c}\right)^2 \frac{mc^2}{\hbar }
(\beta -\beta_0)^2 .
\label{ISPW14}
\end{equation}
The above Eq.(\ref{ISPW14}) for smooth surfaces is in agreement with the 
initial order of magnitude estimate in Eqs.(\ref{est3}) and (\ref{est4}).

\section{Conclusions\label{conc}}

Electromagnetic surface plasma oscillation energies in hydrogen-loaded metal 
cathodes may be combined with the normal electron-proton rest mass energies 
to allow for neutron producing low energy nuclear reactions Eq.(\ref{intro2}). 
The entire process of neutron production near metallic hydride surfaces may 
be understood in terms of the standard model for electroweak interactions. 
The produced neutrons have ultra low momentum since the wavelength is that 
of a low mode isotopic spin wave spanning a surface patch. The radiation energy 
required for such ultra low momentum neutron production may be supplied by the 
applied voltage required to push a strong charged current across the metallic 
hydride cathode surface. Alternatively, low energy nuclear reactions may be induced 
directly by laser radiation energy applied to a cathode surface. They may even be 
induced, albeit at comparatively low rates of reaction, simply by imposing an 
adequate pressure gradient (which can be as little as one atmosphere.) of gaseous 
hydrogen isotopes across a metallic membrane\cite{Iwamura:2002,Li:2003} composed of 
an aggressive hydride-former such as palladium.

The electroweak rates of the resulting ultra low momentum neutron production 
are computed from the above considerations. In terms of the radiation induced 
mass renormalization parameter \begin{math} \beta  \end{math} in Eqs.(\ref{est1}) 
and (\ref{est2}), the predicted neutron production rates per unit area per unit 
time have the form 
\begin{equation}
\varpi_2=\nu (\beta -\beta_0)^2\ \ \ {\rm above\ threshold}
\ \ \ \beta>\beta_0.
\label{conc1}
\end{equation}
The expected range of the parameter \begin{math} \nu \end{math} 
for hydrogen-loaded cathodes is approximately 
\begin{equation}
10^{12}\ \frac{\rm Hz}{\rm cm^2}\  <\ \nu \ < 10^{14}\ \frac{\rm Hz}{\rm cm^2}
\label{conc2}
\end{equation}
in satisfactory agreement with the orders of magnitude observed experimentally. 
More precise theoretical estimates of \begin{math} \nu \end{math} require 
specific material science information about the physical state 
of cathode surfaces which must then be studied in detail. 
As discussed in previous work\cite{Widom:2006}, a deuteron on certain cathode surfaces 
may also capture an electron producing two ultra low momentum neutrons and a neutrino. 
The neutron production rates for heavy water systems are thereby somewhat enhanced.

From a technological perspective, we note that energy must first be put 
into a given metallic hydride system in order to renormalize electron masses and reach 
the critical threshold values at which neutron production can occur.  Net excess energy,  
actually released and observed at the physical device level, is the result of a complex 
interplay between the percentage of total surface area having 
micron-scale E and B field strengths high enough to create neutrons and 
elemental isotopic composition of near-surface target nuclei exposed to local 
fluxes of readily captured ultra low momentum neutrons. In many respects, low 
temperature and pressure low energy nuclear reactions in condensed matter systems resemble 
r- and s-process nucleosynthetic reactions in stars. Lastly, successful fabrication and 
operation of long lasting energy producing devices with high percentages of nuclear active 
surface areas will require nanoscale control over surface composition, 
geometry and local field strengths.

\end{document}